\documentclass[aps,nofootinbib,twocolumn]{revtex4}
\usepackage{bbold}
\usepackage{eufrak}
\usepackage{graphicx}
\usepackage{wasysym}
\usepackage{ulem}
\usepackage[usenames]{color}
\definecolor{DarkGreen}{rgb}{0.0,0.5,0.0}

\begin{document}
\title{Limits on primordial power spectrum resolution: An
inflationary flow analysis}
\author{Brian A. Powell} \email{bapowell@buffalo.edu}
\author{William H.\ Kinney} \email{whkinney@buffalo.edu}

\affiliation{Dept. of Physics, University at Buffalo,
        the State University of New York, Buffalo, NY 14260-1500}
\date{\today}
\begin{abstract}
\noindent
We investigate the ability of current CMB data to reliably
constrain the form of the primordial power spectrum generated during inflation.  We attempt to
identify more exotic power spectra that yield equally good fits to the
data as simple power-law spectra.  In order to test a wide variety of
spectral shapes, we combine the flow
formalism, which is a method of stochastic model generation, with a
numerical integration of the mode equations of quantum
fluctuations.  This allows us to handle inflation models that yield
spectra that
 are not well described by the
standard spectral parameterization.  Using the latest WMAP data-set, we
find a high degree of variation in possible spectral shapes.  In particular, we find
strongly running spectra arising from fast-rolling inflaton fields
providing equally good fits to the data as power-law spectra arising from
slowly-rolling fields.  Current data poorly constrains the spectrum on
scales \(k < 0.01 \ h{\rm Mpc}^{-1}\), where the error due to cosmic
variance is large.  Among the statistically degenerate models, we identify spectra with strong running on these larger scales, but with reduced running at
smaller scales.  These  
models predict values for the tensor-to-scalar ratio, \(r\), that lie outside the 2-\(\sigma\) confidence interval obtained from SDSS+WMAP data for spectra that are parametrized
as power-laws or spectra with constant running.  By considering more generalized power spectra, we therefore open up
regions of parameter space excluded for simpler models.
\end{abstract}
\maketitle

\section{Introduction}
Understanding the nature of the cosmological density perturbation is
essential for the development of inflationary cosmology.  Recent years
have seen a surge in the development of precision cosmology, with
improved measurements of the temperature and polarization anisotropies of
the cosmic microwave background (CMB) as well as increasingly accurate
large scale structure (LSS) surveys.  The task of faithfully reconstructing the physics of
inflation from this data is the
central goal of inflationary phenomenology.  The fundamental physics
responsible for inflation remains elusive, however. Models of inflation
abound with motivations stemming from supergravity, the string landscape,
D-branes, extensions to the standard model and others \cite{Lyth:1998xn,Quevedo:2002xw,Cline:2006hu}.  
A reliable reconstruction of the inflaton
potential, and therefore the physics responsible for inflation, hinges on
accurate measurement and subsequent analysis of cosmological data. There
has been much progress in
the development of methods of primordial power
spectrum reconstruction
\cite{Matsumiya:2001xj,Bridle:2003sa,Mukherjee:2003ag,Hu:2003vp,Hannestad:2003zs,Habib:2004kc,Tocchini-Valentini:2005ja,Shafieloo:2003gf,Shafieloo:2006hs,Bridges:2005br,Bridges:2006zm,Ringeval:2007am}, as well as methods for reconstructing the
inflaton potential
\cite{Lidsey:1995np,Grivell:1999wc,Easther:2002rw,Habib:2004hd,Joy:2005ep,Kadota:2005hv}.  
With the recent release of the WMAP3 data-set \cite{Spergel:2006hy} and with future CMB
missions eminent \cite{planck,inflationprobe}, it is both important and timely to investigate how well current and future experiments will
be able to constrain the form of the primordial power spectrum.  

Recent analyses report a nearly scale invariant spectrum of adiabatic,
gaussian density
perturbations, consistent with the simplest models of single field slow-roll
inflation \cite{Spergel:2006hy,Alabidi:2006qa,deVega:2006hb,Kinney:2006qm,Martin:2006rs,Liddle:2006tc}.  However, experimental and theoretical inaccuracy, in addition to the
limited detection of the polarization signal, limit parameter estimation efforts. 
The dominant source of error is that due to cosmic variance,  a statistical
uncertainty related to the fact that there is only one universe to observe.
This effect is present in both temperature and polarization spectra and is most pronounced on large scales (at low CMB multipoles). 
Cosmic variance results in a fundamental limit on the resolution of CMB spectra, which restricts the accuracy with
which we can reconstruct the physics of the early universe.  Since the primordial
perturbation spectrum provides one of the few pieces of observational evidence for inflation, these uncertainties impose a limit on
how well we can reconstruct the physics of the inflationary era.  While single field slow-roll inflation is well
supported by current data, it is of interest to determine whether more exotic inflationary scenarios might exist within
the resolution limit of observational data.   

In this paper, we study our current ability to constrain the form of the power spectrum.
Using the latest WMAP data-set \cite{Spergel:2006hy}, we generate a best-fit base cosmological
model characterized by 7 parameters, (\(\Omega_bh^2,\Omega_ch^2,h,\tau,A,n_s,r\)).  
We are interested in identifying alternative models with more general power spectra that yield comparable fits to
current data. 
Such statistically degenerate models can be considered effectively indistinguishable, and differences in the  underlying inflation
models unresolvable.  In order to test a wide range of spectral shapes, 
we introduce a Monte Carlo reconstruction program that combines the
inflationary flow technique \cite{Hoffman:2000ue,Kinney:2002qn,Chongchitnan:2005pf} with a numerical integration of the mode equations of quantum fluctuations. Aside from being a novel
reconstruction technique in its own right, we find it highly suitable for
the present investigation. 
We use this method to stochastically generate an ensemble of power spectra to
6\(^{th}\)-order in the flow space.  
For each spectrum it is possible reconstruct the exact inflaton
potential. Because we solve for the spectrum numerically, we can
successfully reconstruct spectra arising from slow-roll as well as strongly non-slow-roll
inflation models.  Additionally, the flow approach allows us to enforce
that each model yield the appropriate amount of inflation, \(N \in
[46,60]\).  

For each spectrum, we fix the non-spectral parameters at the best-fit values and calculate the new model's
likelihood.  We establish a resolution criterion based on the relative likelihoods of the trial models and the
best-fit model.  By fixing the non-spectral parameters, we are not performing a Bayesian analysis of the enlarged parameter space of flow parameters.
By allowing all the parameters to vary as is done in Bayesian parameter estimation, it is
possible that new parameter degeneracies would allow for a wider range of possible spectral shapes to be in agreement with
the data (see, for example, Ref. \cite{Kinney:2001js}). Our method therefore only samples a subset of possible spectra and should be viewed as a more
conservative approach.\footnote{A Bayesian
parameter estimation analysis was recently carried out in \cite{Lesgourgues:2007gp} using a different and slightly smaller parameterization 
than ours.  It is claimed in Ref. \cite{Lesgourgues:2007gp} that the addition of higher-order spectral parameters does not introduce
new parameter degeneracies, and so the results of a Bayesian analysis may be expected to yield similar results to
those obtained here.}
This study is more akin to a model selection analysis: we attempt to identify models constructed from the inflationary
flow space that are equally consistent with the data as models parameterized as
simple power-laws. We utilize the
p-value calculated from the model's chi-squared per degrees of freedom as a measure of statistical significance.  
This approach is frequentist in nature, yet we find general agreement with the
Bayesian study of \cite{Lesgourgues:2007gp}.

This paper is organized as follows:  in section II we review the
inflationary flow formalism and in section III we introduce the new method
which combines the flow approach with a numerical integration of the mode equation.
In 
section IV we investigate primordial power spectrum resolution with 
the WMAP3 data-set.  We find that with TT/TE/EE data alone, our ability to reconstruct the inflaton potential is strongly affected by the
limited spectral resolution.  We find that strongly running spectra arising from non-slow-roll inflation models are currently indistinguishable
from a best-fit power-law model typical of slow-roll inflation.  However, since these non-slow-roll models are
typically accompanied by large tensor components, any information regarding the B-mode polarization signal will
significantly improve resolvability.  Furthermore, we find that the established confidence limits on the
tensor-to-scalar ratio obtained from SDSS and WMAP3 when running is included are violated by the non-power-law
spectra identified in this study.  We find spectra with values for \(r\) at \(k = 0.002 \ h{\rm Mpc}^{-1}\) that lie outside
the 2-\(\sigma\) contour of the \(n_s\)-\(r\) marginalized likelihood \cite{Spergel:2006hy,Kinney:2006qm} yield {\it equally good} fits to the data
as the best-fit power-law. 
Section V includes discussion and conclusions.  

\section{Single field inflation and The Flow Formalism}

The evolution of a Friedmann-Robertson-Walker (FRW) universe dominated by a single minimally-coupled scalar field (the
{\it inflaton}) with potential \(V(\phi)\) is given by the equations
\begin{eqnarray}
\label{fe1}
H^2 &=& \frac{8\pi}{3m_{\rm Pl}^2}\left[\frac{1}{2}\dot{\phi}^2 + V(\phi)\right],  \\
\label{fe2}
\frac{\ddot{a}}{a} &=& \frac{8\pi}{3m_{\rm Pl}^2}\left[V(\phi) - \dot{\phi}^2\right].
\end{eqnarray}
We have assumed a flat universe described by the metric \(g_{\mu \nu} = {\rm diag}(1,-a^2,-a^2,-a^2)\), where
\(a(t)\) is the scale factor and dots denote derivatives with respect to coordinate time, \(t\).
The evolution of the scalar field follows from stress-energy conservation and takes the form of a Klein-Gordon equation with a
cosmological friction term,
\begin{equation}
\label{kg}
\ddot{\phi} + 3H\dot{\phi} + V' = 0,
\end{equation}
where primes denote derivatives with respect to the field, \(\phi\).  Equations (\ref{fe1}) and (\ref{kg}) can be combined to yield
the alternative equations of motion,
\begin{eqnarray}
\label{hje}
\dot{\phi} &=& -\frac{m^2_{\rm Pl}}{4\pi}H'(\phi), \\
H'(\phi)^2 - \frac{12\pi}{m^2_{\rm Pl}}H^2(\phi) &=& -\frac{32{\pi}^2}{m^4_{\rm Pl}} V(\phi), 
\end{eqnarray}
where the Hubble parameter, written as a function of \(\phi\), becomes the dynamical variable.  The field \(\phi\) serves as a
convenient time variable so long as it is monotonic.  The second of the above two equations is known as the {\it Hamilton-Jacobi
equation}, and may be written more simply as 
\begin{equation}
H^2(\phi)\left[1-\frac{1}{3}\epsilon(\phi)\right] =
\left(\frac{8\pi}{3m_{\rm Pl}^2}\right)V(\phi),
\end{equation}
where the parameter \(\epsilon\) is defined as
\begin{equation}
\label{epsilon}
\epsilon =
\frac{m^{2}_{{\rm Pl}}}{4\pi}\left(\frac{H'(\phi)}{H(\phi)}\right)^2.
\end{equation}
Physically, \(\epsilon\) is the equation-of-state parameter of the cosmological fluid, and from Eq. (\ref{fe2}) it follows that
the condition for inflation, \(\ddot{a} > 0\), requires that \(\epsilon < 1\).  

Starting with the equation-of-state parameter,
it is possible to define an infinite hierarchy of parameters \cite{Liddle:1994dx} by taking successive
derivatives of the Hubble parameter, \(H(\phi)\),
\begin{eqnarray}
\label{flowpar}
\eta &=& \frac{m^{2}_{{\rm Pl}}}{4\pi}\left(\frac{H''(\phi)}{H(\phi)}\right),\\ \nonumber
\xi^2 &=&
\frac{m^{4}_{{\rm Pl}}}{(4\pi)^2}\left(\frac{H'(\phi)H'''(\phi)}{H^2(\phi)}\right),\\ \nonumber
\vdots \\ \nonumber
^n\lambda_H &=&
\left(\frac{m^{2}_{{\rm Pl}}}{4\pi}\right)^n\frac{(H'(\phi))^{n-1}}{H^n(\phi)}\frac{d^
{(n+1)}H}{d\phi^{(n+1)}}. 
\end{eqnarray} 
In what follows, we will refer to
Eqs. (\ref{epsilon}) and (\ref{flowpar}) collectively as the {\it Hubble flow
parameters}.  These are often referred to as {\it slow-roll} parameters in the literature, however, they are defined here without any
assumption of slow-roll.  For this analysis it is most convenient to use the number of e-folds
before the end of inflation, \(N\), as our time variable.  The scale factor may be written \(a \propto e^{N}\) and from Eq.
(\ref{fe1}),
\begin{equation}
\label{N}
dN = -H dt = \frac{H}{\dot{\phi}}d\phi = \frac{2\sqrt{\pi}}{m_{\rm Pl}}\frac{d\phi}{\sqrt{\epsilon(\phi)}}.
\end{equation}
Making use of this relation, we take successive derivatives of the flow parameters with respect to \(N\),  generating an infinite set of differential equations
\cite{Kinney:2002qn},
\begin{eqnarray}
\label{floweqs}
\frac{dH}{dN} &=& \epsilon H,   \\
\frac{d\epsilon}{dN} &=& \epsilon(\sigma + 2\epsilon),   \nonumber \\
\frac{d\sigma}{dN} &=& -5\epsilon\sigma - 12\epsilon^2 + 2\xi^2, \nonumber \\
\frac{d(^{\ell}\lambda_H)}{dN} &=& \left[\frac{\ell - 1}{2}\sigma + (\ell -
2)\epsilon\right](^{\ell}\lambda_H) + {^{\ell + 1}}\lambda_H, \nonumber
\end{eqnarray}
where \(\sigma = 2\eta - 4\epsilon\).  In practice, this system is truncated at some
finite order \(M\) by requiring that \(^{M+1}\lambda_H = 0\).  This system can then be
solved numerically by specifying the initial conditions of the parameters \(\epsilon\),
\(\sigma, \dots,  {^{M}\lambda_H}\) at some arbitrary time, \(N_i\).  Although the system is truncated at finite order, this
results in an {\it exact} solution for the background evolution of an FRW
universe dominated by a single scalar field. This is due to the form of the flow equations, where it can be seen that the truncation \(^{M+1}\lambda_H = 0\)
ensures that all higher-order parameters vanish for all time. 
In Ref. \cite{Kinney:2002qn}, the initial conditions were drawn randomly from the ranges
\begin{eqnarray}
\label{ics}
\epsilon &\in& [0,0.8] \\
\sigma &\in& [-0.5,0.5]  \nonumber \\
\xi^2 &\in& [-0.05,0.05]  \nonumber \\
^3\lambda_H &\in& [-0.005,0.005]  \nonumber \\
\vdots  \nonumber \\
^{M+1}\lambda_H &\in& 0, \nonumber
\end{eqnarray}
although other choices are possible.  
The system is then
evolved forward in time until either inflation ends (\(\epsilon > 1\)), or the system
reaches a late-time asymptote, \(\epsilon \rightarrow 0\).  The latter possibility arises
in models in which the field evolves to a point of nonzero vacuum energy, leading to
eternal inflation.  In such cases, inflation must end via the action of an auxiliary
field, such as in hybrid models \cite{Linde:1993cn}.  Such models generically
predict scalar spectra with \(n_s > 1\) and negligible running, in
conflict with the WMAP3 data-set \cite{Spergel:2006hy}.  We therefore focus solely on models for which inflation
ends through a failure of slow-roll, \(\epsilon > 1\), originally termed {\it nontrivial} models \cite{Kinney:2002qn}.  

Once a nontrivial model is found, the flow equations are integrated
backwards in time from \(N = 0\) to \(N = N_{obs}\), where \(N_{obs}\) is drawn randomly
from the range \([40,60]\).  The solution to the flow equations then comprise the full time
evolution of the parameters \(\epsilon\), \(\sigma, \dots,
{^{M}\lambda_H}\) from \(N = 0 \rightarrow N_{obs}\).  It is then
possible to reconstruct the inflaton potential using the Hamilton-Jacobi
equation, Eq. (\ref{hje}) \cite{Easther:2002rw}.  The values of the flow parameters at
\(N_{obs}\) can then be
used to calculate observables at this point via the relations \cite{Stewart:1993bc}
\begin{eqnarray}
\label{obs}
r &=& 16\epsilon[1-C(\sigma+2\epsilon)], \\
n_s &=& 1+ \sigma - (5-3C)\epsilon^2  \nonumber \\
 && -\frac{1}{4}(3-5C)\sigma\epsilon+\frac{1}{2}(3-C)\xi^2,  \nonumber \\
\alpha &=& \frac{dn_s}{d{\rm ln}k} = -\left(\frac{1}{1-\epsilon}\right)\frac{dn_s}{dN}, \nonumber
\end{eqnarray}
where \(C = 4({\rm ln}2+\gamma)-5\), \(\gamma \simeq 0.577\), and \(\alpha\) denotes the running of the scalar spectral index, \(n_s\).
By adopting a Monte Carlo approach, large numbers of models can be generated and their
observable predictions compared with current observational data.  This serves as a means for
constructing classes of models that satisfy certain criteria, as well as gaining insight
into the generic features of the inflationary parameter space.
Models generated stochastically using the above method are observed to cluster strongly in
the \(n_s-r\) and \(\alpha-n_s\) planes
\cite{Hoffman:2000ue,Kinney:2002qn,Calcagni:2005hy}.
It is difficult, however,  to attribute any rigorous statistical meaning to such clustering, since
there is no well-defined measure on the parameter space. 

The main drawback of the flow method is that while generating exact inflationary evolutions, it must rely on approximations when
calculating the observable predictions of these models. 
These approximations are made as two separate series truncations.
The first approximation has to do with the parameterization of the power spectrum.  When quoting observables in terms of \(r\), \(n_s\) and \(\alpha\), one is considering a
truncated Taylor expansion
of \({\rm ln}(P(k))\) in \({\rm ln}(k)\),
\begin{equation}
\label{spec}
{\rm ln}\left(\frac{P(k)}{P(k_0)}\right) = (n_s - 1){\rm ln}\left(\frac{k}{k_0}\right) + \frac{1}{2}\alpha {\rm
ln}\left(\frac{k}{k_0}\right)^2 + \cdots,
\end{equation}
where \(k\) is the comoving wavenumber.  
Clearly, this parameterization is only useful if the higher-order terms are small across the range of scales for which it is expected to
hold.  The second approximation results from the use of Eq. (\ref{obs}) to connect these spectral parameters to the inflationary flow
parameters.
These expressions are only accurate to order \(\mathcal{O}(\epsilon^2)\) in the
flow parameters.  While the evolution of the lowest-order parameters appearing explicitly in Eq.
(\ref{obs}) are determined by the full set of flow parameters out to order \(M\), the use of
these expressions still requires that these higher-order parameters be negligible on observable
scales.  While, in principle, Eq. (\ref{obs}) can be extended to
arbitrary order in slow-roll,
the expressions quickly become very algebraically complex \cite{Gong:2001he,Choe:2004zg}.  Furthermore, this approach evaluates observables at `horizon crossing', a technique that is not always applicable \cite{Wang:1997cw,Leach:2001zf,Kinney:2005vj}.

These approximations are perfectly valid when considering inflation
models that satisfy the slow-roll criteria, \(\epsilon\), \(\eta \ll 1\).  In this case Eq. (\ref{obs}) is highly accurate and
higher-order spectral parameters are negligible, validating the use of Eq. (\ref{spec}).  However, if one wishes to use flow
methods to explore regions of the inflationary parameter not well approximated by slow-roll, this approach must be adapted.
As pointed out by Easther and Peiris \cite{Peiris:2006ug}, if one imposes an
inflationary prior on the underlying model space, then there is no need to calculate observables in terms of the
spectral parameters.   The flow parameters
themselves fully specify the physics of the inflationary epoch and one avoids introducing
further, unnecessary errors.   We adopt this philosophy in what follows, doing away
completely with the parameterization Eq. (\ref{spec}) and the specification of observables Eq. (\ref{obs}). 

We develop an improved method of reconstruction by combining the flow formalism with a
numerical evaluation of the mode equation of inflationary perturbations.  By solving the mode equation for multiple
\(k\)-values, it is possible to reconstruct the primordial power spectra without recourse to
spectral parameters or the  slow-roll approximation.
This leads to a consistent match between
the scalar field potential and associated perturbation spectra, to the same degree of
accuracy.  In addition, it allows for an investigation of regions of the inflationary parameter
space that lie far from slow-roll, leading to a more robust reconstruction program. (For other efforts at numerical evaluation of inflationary perturbations, see Refs. \cite{Salopek:1988qh,Adams:2001vc,Martin:2006rs,Ringeval:2007am}.)

\section{Calculating the primordial power spectrum}  
During inflation, vacuum fluctuations of the inflaton field are redshifted to
superhorizon scales by the rapidly expanding spacetime where they become classical
curvature perturbations.  The inflaton field couples at linear order to the scalar metric
perturbation, \(\delta g^s_{\mu \nu}\), which may be defined in terms of four scalar
functions,
\begin{eqnarray}
\delta g^s_{00} &=& 2A \nonumber \\
\delta g^s_{0i} &=& \partial_i B \nonumber \\
\delta g^s_{ij} &=& 2(\mathcal{H}_L \delta_{ij} + \partial_i \partial_j \mathcal{H}_T). 
\end{eqnarray}
If one works in {\it comoving gauge}, in which the spatial slices are orthogonal to the
worldlines of
comoving observers, the intrinsic curvature perturbation of the spacelike hypersurface is
\cite{Bardeen:1980kt,Liddle:2000cg}
\begin{equation}
\label{R}
\mathcal{R} = \mathcal{H}_L + \frac{1}{3}\mathcal{H}_T.
\end{equation}
In comoving gauge, one has the additional freedom of requiring that the spatial part of the
metric perturbation be isotropic, \(\mathcal{H}_T = 0\), so that \(\mathcal{R} = \mathcal{H}_L\).  
The coupling between inflaton and metric perturbations motivates the use of the gauge invariant Mukhanov potential \cite{Mukhanov:1990me,Mukhanov:1988jd,Mukhanov:1985rz},
\begin{equation}
u = a\delta\phi - \frac{{\phi'}}{H}\mathcal{H}_L,
\end{equation}  
where \(a\) is the scale factor of the universe, \(H\) is the Hubble parameter and
\(\delta \phi\) is the scalar field fluctuation.  Primes denote derivatives with respect to conformal time, \(\tau\).  
On comoving hypersurfaces, \(\delta \phi = 0\), and the Mukhanov potential is related to the comoving curvature perturbation,
\begin{equation}
\mathcal{R} = \left|\frac{u}{z}\right|,
\end{equation}
where \(z = \phi'/H\).  
The two-point correlation function can be written in terms of the power spectrum, \(P_{\mathcal R}(k)\),
\begin{equation}
\langle \mathcal{R}_{\bf k} \mathcal{R}_{\bf k'}\rangle = \frac{2 \pi^2}{k^3}P_{\mathcal R}(k)\delta({\bf k} - {\bf k'}).
\end{equation}
The power spectrum of the curvature perturbation may then be
written,
\begin{equation}
\label{specs}
P_{\mathcal{R}}(k) = \frac{k^3}{2\pi^2}\left|\frac{u_k}{z}\right|^2,
\end{equation}
where the \(u_k\) are the Fourier modes of the gauge-invariant potential satisfying
the equation of motion,
\begin{equation}
\label{modeq}
u_k'' + \left(k^2 - \frac{z''}{z}\right)u_k = 0.
\end{equation}

Tensor metric perturbations, \(\delta g^T_{\mu \nu}\), are also excited during inflation, leading to a large-scale
gravitational wave background.   The metric perturbation is purely spatial and may be
described by a single function, \(\delta g^T_{ij} = h_{ij}\).  The dynamics of these
perturbations are determined by the linearized Einstein equations, which follow from the
action \cite{Grishchuk:1974ny},
\begin{equation}
\label{gw}
S_h = \frac{m^2_{\rm Pl}}{64 \pi} \int d\tau d^3{\bf x} \, a^2(\tau) \partial_\mu h^i_{\ j} \partial_\nu h_i^{\ j}.
\end{equation}
Because gravitational waves are both transverse and traceless ({\it i.e.} \(h^i_{\ i} = \partial_i h_{ij} = 0\)), they may be
decomposed into two independent polarization modes, denoted + and \(\times\).  The Fourier decomposition may then be written
\begin{eqnarray}
&&h_{ij}(x) = \\
&&\int \frac{d^3{\bf k}}{(2\pi)^{3/2}}\left[h_{{\bf k},+}(\tau)e^+_{ij}({\bf k}) + h_{{\bf k},\times}(\tau)e^\times_{ij}({\bf
k})\right]e^{i{\bf k \cdot x}}, \nonumber
\end{eqnarray}
where we have used the fact that \(h^*_{\ {\bf k}} = h_{\bf -k}\) which follows from the condition \(e^*_{ij}({\bf k}) =
e_{ij}(-{\bf k})\).  The power spectrum of tensor fluctuations can then be be written
\begin{equation}
\label{spech}
P_h(k) = \frac{k^3}{2\pi^2}\left(\langle|h_{{\bf k},+}|^2\rangle + \langle|h_{{\bf k},\times}|^2\rangle\right).
\end{equation}
Following the field redefinition,
\begin{equation}
v_{+,\times} = \sqrt{\frac{a^2 m^2_{\rm Pl}}{32 \pi}}h_{+,\times},  
\end{equation}
the equation of motion that follows from Eq. (\ref{gw}) for each polarization mode becomes that of a canonically normalized
massless scalar field in an FRW
background,
\begin{equation}
\label{tensoreq}
v_k'' + \left(k^2 - \frac{a''}{a}\right)v_k = 0,
\end{equation}
where the polarization indices have been suppressed. 
Here \(v_k\) is the quantum mode associated with positive-frequency excitations of the metric perturbation.  
In terms of the new field, the power spectrum Eq. (\ref{spech}) becomes
\begin{equation}
\label{tspec}
P_h(k) = \frac{32 k^3}{\pi}\left|\frac{v_k}{a}\right|^2.
\end{equation}
Although written in terms of a single solution of Eq. (\ref{tensoreq}), this expression includes the contributions from both
polarization modes.

Our goal is to combine Eqs. (\ref{modeq}) and (\ref{tensoreq}) with the flow equations, Eq. (\ref{floweqs}).  Then, for any
initial point in the flow space, we will be able to determine the full evolution of the modes \(u_k\) and \(v_k\)
together with the background evolution.  Since the most convenient time variable for use with the flow formalism is \(N\), the
number of e-folds before the end of inflation, it is necessary to recast Eqs. (\ref{modeq}) and (\ref{tensoreq}) in terms of
\(N\).  These expressions become
\begin{equation}
\label{modeqN}
\frac{d^2u_k}{dN^2} + (\epsilon - 1)\frac{du_k}{dN} +
\left[\left(\frac{k}{aH}\right)^2 - F(\epsilon,\sigma,\xi^2)\right]u_k = 0,
\end{equation}
and
\begin{equation}
\label{tensoreqN}
\frac{d^2v_k}{dN^2} + (\epsilon - 1)\frac{dv_k}{dN} +
\left[\left(\frac{k}{aH}\right)^2 - (2 - \epsilon)\right]v_k = 0,
\end{equation}
where the function \(F(\epsilon,\sigma,\xi^2)\) of Eq. (\ref{modeqN}) is defined as
\begin{equation}
F(\epsilon,\sigma,\xi^2) = 2\left(1 - 2\epsilon - \frac{3}{4}\sigma - \epsilon^2 +
\frac{1}{8}\sigma^2 + \frac{1}{2}\xi^2\right).
\end{equation}
The full system of differential equations is formed by Eqs. (\ref{modeqN}) and
(\ref{tensoreqN}) together with the flow equations, Eq. (\ref{floweqs}).  
The standard choice of initial conditions for the mode function is that defined
by the Bunch-Davies vacuum,
\begin{equation}
\label{bd}
u_k(-k\tau \rightarrow \infty) = \sqrt{\frac{1}{2k}}e^{-ik\tau}.
\end{equation}
When solving the mode equations numerically we cannot use this exact condition for two reasons.
First, we cannot set this condition in the infinite past, but must impose it at a finite time.    
Imposing the limit Eq. (\ref{bd}) at a finite time, {\it i.e.} finite length scale, results in
modulations of the power spectrum akin to those arising from
transplanckian effects \cite{Easther:2002xe}.
This effect can be minimized by initializing the mode functions at sufficiently early times (small
length scales). 
Second, we must write Eq. (\ref{bd}) in terms of \(\tau(N)\) for use with Eqs (\ref{modeqN}) and (\ref{tensoreqN}), and this function is not known in
general.  From the relation \(dN = -aHd\tau\), we obtain
\begin{equation}
\frac{dy}{d\tau} =-k(\epsilon -1),
\end{equation}
where \(y = k/aH\) is the ratio of the Hubble radius to the proper wavelength of the
fluctuation and is a function of \(N\).
If \(\epsilon(y)\) is approximately constant, this equation can be integrated to give
\(\tau(y)\).  From the equation
\begin{equation}
\frac{d\epsilon}{dy} = \frac{1}{y(1-\epsilon)}\frac{d\epsilon}{dN},
\end{equation}  
we see that \(\epsilon(y) = const.\) if \(y\) is taken sufficiently large.  By taking \(y\)
large, we are also ensuring that the modes are initialized in the short-wavelength limit.  
This motivates the use of the approximate initial conditions,
\begin{eqnarray}
\label{cond}
u_{k}(y_i) &=& \sqrt{\frac{1}{2k}}e^{-iy_i/(1-\epsilon_i)} \nonumber \\
\left.\frac{du_k}{dN}\right|_{y=y_i} &=& \sqrt{\frac{1}{2k}}y_ie^{-iy_i/(1-\epsilon_i)},
\end{eqnarray}
For a choice of initial flow
parameters \(\epsilon_i\), \(\sigma_i\), ..., \(^{M-1}\lambda_{Hi}\), Eq.
(\ref{ics}), we set the initial condition for
each \(k\)-mode at \(y_i/(1-\epsilon_i) = 100\).  This proves to be sufficiently
large to ensure the accuracy of the conditions Eq. (\ref{cond}). 

Since there are two complex solutions to each of Eqs. (\ref{modeqN}) and (\ref{tensoreqN}),
rather than work with complex coefficients, we define the orthogonal solution basis,
\begin{eqnarray}
u_{k,1} &=& \frac{u_k + u_k^*}{2} \nonumber \\ 
u_{k,2} &=& \frac{u_k - u_k^*}{2i}. 
\end{eqnarray}
Each mode is evolved from \(N(y_i/(1-\epsilon_i)=100)\) to \(N = 0\), the end of
inflation.  The amplitude of the power spectrum for each \(k\)-mode is
then obtained by evaluating Eqs. (\ref{specs}) and (\ref{tspec}) at this time, when all modes have attained the long-wavelength limit.

\section{Spectrum resolution with WMAP3}
In this section we investigate the ability of the latest WMAP data-set to accurately resolve the form of the power spectrum.
We approach this problem by considering a 7-parameter fiducial best-fit model parameterized by
inflationary and non-inflationary degrees of freedom.  We suppose that this model accurately
describes the universe and search for alternative models which are statistically
indistinguishable from it.
The non-inflationary parameters are taken to be the baryon
and CDM densities, \(\Omega_b h^2\) and \(\Omega_c h^2\), the Hubble parameter, \(h\), the optical depth to reionization, 
\(\tau\),  and the overall spectrum normalization, \(A(k = 0.002 \, h
{\rm Mpc}^{-1})\).\footnote[2]{Since the overall spectrum amplitude is not fixed by
the inflationary model in the flow formalism, it is considered a non-inflationary parameter in this
analysis.}  The inflationary parameters are the scalar spectral index, \(n_s\), and
the tensor contribution, \(r\).  We consider purely adiabatic initial perturbations and assume spatial
flatness. The tensor spectral index, \(n_t\), is assumed to satisfy the
inflationary consistency condition, \(n_t = -r/8\), and does not represent an additional free
parameter.  

Since we only vary the form of the spectrum in this analysis, we fix the non-inflationary parameters at their best-fit values.
We then replace the power-law parameterization (\(n_s\), \(r\)) with tensor and scalar power spectra generated
with
the Monte Carlo method.  We work to \(6^{th}\)-order in the flow space,
{\it i.e.} \(^i \lambda_H = 0\) for \(i \geq 6\).  
We use a version of \texttt{CAMB} \cite{Lewis:1999bs} modified to accept arbitrary power spectra as input to generate the associated
\(C_\ell\)-spectra.  It is then possible to calculate the model's effective chi-square, \(\chi^2_{eff} = -2{\rm
ln}{\mathcal L}\), using the WMAP3 likelihood software available at the \texttt{LAMBDA}
website\footnote[3]{http://lambda.gsfc.nasa.gov/}.  One can then collect models that satisfy specific likelihood
criteria relative to the best-fit model.  

How does one compare the statistical significance of two different models?  If
the two models have the same parameterization, then the confidence limits
obtained from a maximum likelihood analysis of the parameter space suffice.  However, in
this study, we wish to  compare a model comprising 2 spectral parameters with a model
comprising 6 flow parameters.  A simple statistic used in model selection
analyses is the likelihood ratio test, \(2{\rm ln}\mathcal{L_{\rm
simple}}/\mathcal{L_{\rm complex}}\), where the simple model is so named because
it contains fewer free parameters than the more complex model.  This statistic
is approximately \(\chi_\nu^2\) distributed with \(\nu\) degrees of degrees of
freedom equal to the difference in the number of free parameters
between the complex and the simple model.  The significance level associated
with the value of the \(\chi^2_\nu\) can then be used to quantify the relative
goodness-of-fit between these models.  Use of this test requires that the two
models be {\it nested} - that the complex model is formed by adding parameters
to the base model.  The space of flow parameters and spectral parameters are at
best only approximately nested: there is no way to map the flow 
parameters \((\epsilon,\sigma,\xi^2, ^3\lambda_H, ^4\lambda_H, ^5\lambda_H)\) to a
finite number of spectral parameters.
This is because there are infinitely many higher-order spectral
parameters that are functions of these 6 flow parameters that in general will not
vanish under the mapping.  We say {\it approximate} because if these parameters are
very small, they might be neglected.  A further downside is that it is not clear how to tell
from this statistic if two models are of comparable significance.

Another method often used in model selection analyses is the Bayesian
information criterion (BIC) \cite{schwartz},
\begin{equation}
{\rm BIC} = -2{\rm ln}{\mathcal L} + k{\rm ln}N,
\end{equation}
where \({\mathcal L}\) is the maximum likelihood of the model, \(k\) the number
of free parameters and \(N\) the number of data points.  
The BIC is an approximation of the Bayesian evidence, which is the integral of
the likelihood function over the full parameter space.  
The BIC penalizes overparameterized models that don't provide significantly
better fits to the data.  
This approach is also
ill-suited for our purposes, because it assumes that all \(k\) parameters are
well measured by the data.  In fact, it is the opposite of this case that
motivates this study: current data is {\it not} good enough to resolve higher order
terms in the spectral decomposition, and we seek to determine what these
resolution limits are.   
  
In what follows, we instead consider the p-value calculated from a model's chi-square per degrees of freedom as a measure of
goodness-of-fit.  This method is applicable so long as the likelihoods are
approximately Gaussian.
The p-value of a proposed model with a given \(\chi^2\) is  
\begin{equation}
\label{pval}
p_\nu = \int_{\chi^2}^{\infty} P_\nu(y) \, dy,
\end{equation}
where \(P_\nu\) is the chi-square probability distribution function with \(\nu\) degrees of
freedom (d.o.f.),
\begin{equation}
P_\nu(y) = \frac{\left(\frac{1}{2}\right)^{\nu/2}}{\Gamma(\nu/2)}y^{\nu/2 -1}e^{-y/2}.
\end{equation}
For a given {\it significance level}, \(\alpha\), if \(p < \alpha\) then the proposed model may be
rejected at the \(1-\alpha\) confidence level.  
More precisely, the p-value is the probability that we obtain a
particular \(\chi^2\) due to chance alone. It can therefore be
interpreted as the probability of falsely rejecting a correct
model of the universe.
To determine relative goodness-of-fit between two models, one simply compares
their p-values.  
In order to be as conservative as
possible, we consider trial models lying within \(|\Delta p| \leq 0.01\) of the best-fit model to be effectively
indistinguishable from it.  
The spectra generated by \(6^{th}\)-order Monte Carlo are described by 6 free parameters, corresponding to the flow parameters \(\epsilon\),
\(\sigma\), ..., \({^5}\lambda_H\).  Therefore, the trial spectra contain 4 more parameters
than the best-fit power-law model. 
The difference in p-value between a trial model with \(\nu-4\) d.o.f. and the
best-fit model with \(\nu\) d.o.f.,
\begin{equation}
|\Delta p| = |p_{\nu-4} - p_\nu| \leq 0.01, 
\end{equation}
corresponds to the likelihood  spread
\begin{equation}
-8\times10^{-4}
\lesssim \chi^2_{\rm trial}/(\nu-4) - \chi^2_{\rm best-fit}/\nu \lesssim 10^{-3}.  
\end{equation}
The chi-square per degrees of freedom of the best-fit model is \(\chi^2_{\rm
best-fit}/\nu = 1.0210\).

\begin{figure}
%\centerline{\includegraphics[angle=270,width=4in]{spectra.ps}}
\centerline{\includegraphics[angle=270,width=3.75in]{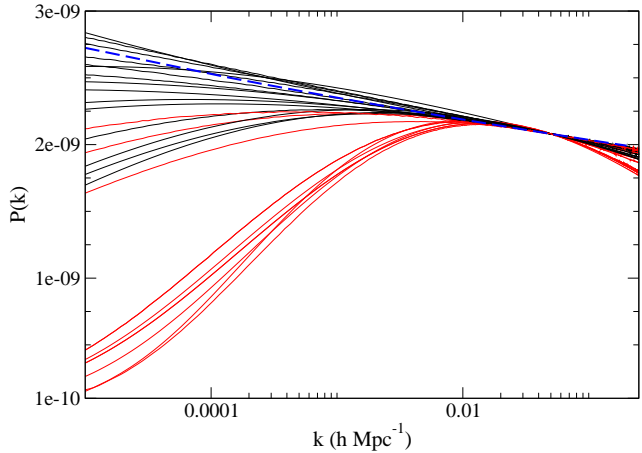}}
\caption{A representative sampling of power spectra lying within \(|\Delta p| \leq 0.01\) of the best-fit model.
The spectrum of the best-fit model is represented by the heavy dashed
line.  The spectra are colored-coded according to their predicted values for \(r\) at \(k = 0.002 \, h{\rm
Mpc}^{-1}\): red yield \(r \sim \mathcal{O}(10^{-1})\) and black yield \(r < \mathcal{O}(10^{-1})\).}
\label{figure1}
\end{figure}
\begin{figure}
%\centerline{\includegraphics[angle=270,width=4in]{spectra.ps}}
\centerline{\includegraphics[angle=270,width=3.75in]{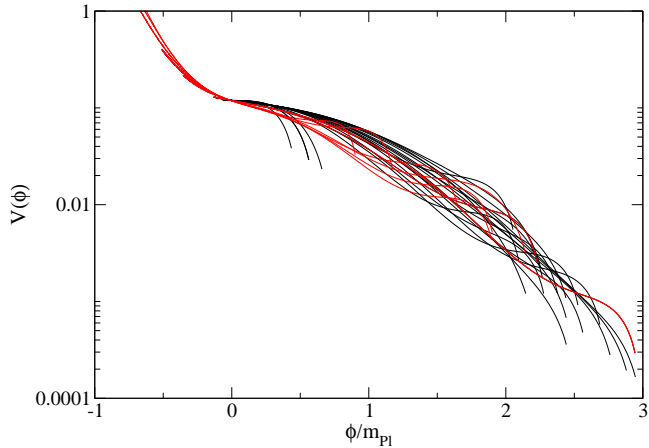}}
\caption{A representative sampling of reconstructed potentials lying within \(|\Delta p| \leq 0.01\) of the best-fit model.
The potentials are colored-coded according to their predicted values for \(r\) at \(k = 0.002 \, h{\rm
Mpc}^{-1}\): red yield \(r \sim \mathcal{O}(10^{-1})\) and black yield \(r < \mathcal{O}(10^{-1})\).  The potentials
have been given a common, arbitrary normalization at \(\phi =0\), when scales
corresponding to the quadrupole
exit the horizon.}
\label{figure5}
\end{figure}

We begin with a best-fit model with \(n_s = 0.969\), \(A = 2.30 \times
10^{-9}\) and a tensor-to-scalar ratio, \(r = 0.0346\).  In Figure
\ref{figure1} we present a sample of power spectra lying within \(|\Delta p| \leq 0.01\) of this model, and in
Figure \ref{figure5} some reconstructed potentials. 
In order to identify spectra exhibiting the strongest deviation from power-law, we initially imposed a non-slow-roll
prior on the model space. In order for a model to be considered for full power spectrum integration, we require
that the spectral index calculated to \(2^{nd}\)-order in slow-roll Eq. (\ref{obs}) differ from the
\(3^{rd}\)-order result \cite{Choe:2004zg} by at least \(1\%\).  This search resulted in the grouping of red-colored spectra
exhibiting a large suppression of power on large scales in Figure \ref{figure1}.  We later relaxed this prior to
obtain the black-colored spectra, allowing us to form a degeneracy envelope. 

\begin{figure}
\centerline{\includegraphics[angle=270,width=3.75in]{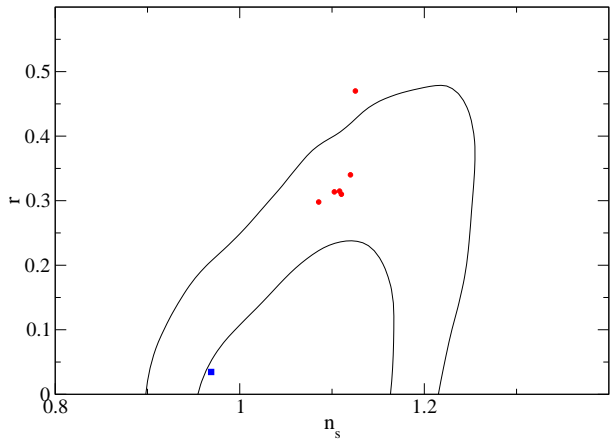}}
\caption{Constraints on the \(n_s\)-\(r\) plane from combined SDSS and WMAP data for spectra with constant running.  The inner
and outer contours mark the \(68 \%\)- and \(95 \%\)-confidence intervals, respectively.  The red points denote
the values of \(n_s\) and \(r\) of the red spectra in Figure \ref{figure1}.  The blue square is that of the
fiducial power-law model.} 
\label{figure4}
\end{figure}

Remarkably, the red spectra remain equally good fits to the data as the fiducial power-law even when SDSS data
\cite{Tegmark:2001jh} is included.  
For the case of power-law spectra, the
latest analyses report \(r_{0.002} < 0.30\) at \(95 \%\)-confidence when SDSS is combined with WMAP3, while \(r_{0.002} < 0.65\) at
\(95 \%\)-confidence with WMAP3 alone \cite{Spergel:2006hy}.  The error bars open up considerably when
running is allowed: \(r_{0.002} < 0.38\) at \(95 \%\)-confidence SDSS+WMAP3, versus 
\(r_{0.002} < 1.1\) with WMAP3 alone.
Spectra with larger \(r_{0.002}\) can be accommodated by the
data if there is suppressed scalar power on these scales.  As
\(r\) decreases on smaller scales, the scalar spectrum must run to larger amplitudes in order to maintain the
correct amount of overall power.  The SDSS data provides an accurate measurement of the matter power
spectrum on scales \(0.01 \ h{\rm Mpc^{-1}} \apprle k \apprle 0.3 \ h{\rm Mpc^{-1}}\), and while not probing scales
\(k \sim 0.002 \ h{\rm Mpc^{-1}}\) directly, if there is a strong negative {\it constant} running of the spectrum on these scales it may lead to a measurable change in power on the
intermediate scales that are directly probed by SDSS.  In particular, if the running is too large at \(k \sim 0.002 \ h{\rm Mpc^{-1}}\), there will be a loss of power
on smaller scales as the spectrum dips blue again.  This results in the much tightened bound on
\(r\).  
While the red spectra in Figure \ref{figure1} are strongly running on
large scales, permitting a sizeable tensor-to-scalar ratio, this running is {\it not constant} and is unsubstantial on
small scales.  Indeed, when the red spectra are plotted in the \(n_s\)-\(r\) plane, we find that they all
lie outside the 1-\(\sigma\) contour of the SDSS+WMAP marginalized likelihood, Figure \ref{figure4}.  
Therefore, by considering more general spectra, the error bars inevitably get much larger.  
The spectral index of the non-power-law spectra is 
calculated from the slope of the spectrum at \(k = 0.002 \ h{\rm Mpc^{-1}}\),
\begin{equation}
n_s = \left. \frac{d{\rm ln}P(k)}{d{\rm ln}k}\right|_{k = 0.002}.
\end{equation}
This is a measure of the local value of \(n_s\) around \(k = 0.002 \ h{\rm Mpc^{-1}}\) and it should
be emphasized that the large uncertainty is due to the generality of the spectral shapes.  However, the
uncertainty in the local value of \(n_s\) is drastically reduced at smaller scales, as indicated in Figure
\ref{figure1}.  Recently, Cortes {\it et al.} \cite{Cortes:2007ak} determined that the effect of the degeneracy in the
\(n_s\)-\(r\) plane can be minimized by quoting observables at \(k = 0.017 \ h{\rm Mpc^{-1}}\).  

\begin{figure}
\centerline{\includegraphics[angle=270,width=3.75in]{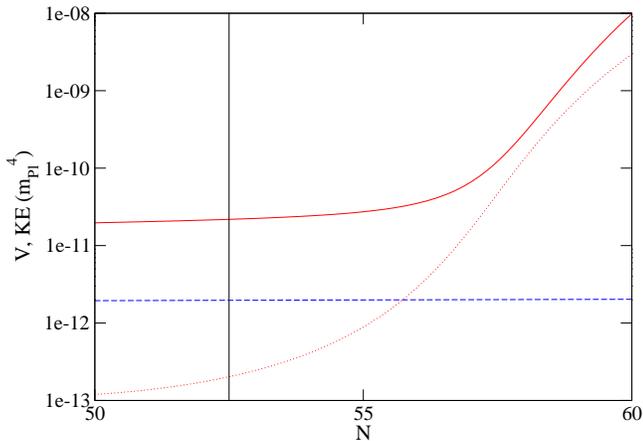}}
\caption{The best-fit inflaton potential (blue dashed line) as compared to a model lying
within \(\Delta p \leq 0.01\) of this model.  The potential energy of the
alternative model is
given by the red solid line, and the kinetic energy is given by the red dotted line.  The
vertical line denotes the time when scales corresponding to the quadrupole leave the horizon.}
\label{figure2}
\end{figure}
The statistical degeneracy of the power spectra exhibited in Figure \ref{figure1} translates directly to a degeneracy in the space of inflation
models, Figure \ref{figure5}.
The red-colored inflation models that give rise to the red-colored non-power-law
spectra are characterized by an initially {\it fast-rolling} inflaton that has failed to reach the slow-roll attractor by
the time observable scales exit the horizon, similar to models proposed by Contaldi, {\it et al.} \cite{Contaldi:2003zv}.  In Figure \ref{figure2} we plot one such potential.  The kinetic energy
is initially large and  monotonically
decreasing with respect to time.  
From the equation of motion Eq. (\ref{kg}), it is evident that if the field velocity is initially large, the Hubble drag term \(3H\dot{\phi}\) dominates the
evolution, slowing the field.
The field decelerates until the drag term becomes subdominant near the
end of inflation.  In contrast, the kinetic energy of the best-fit power-law model (not included in the
figure) is \({\mathcal O}(10^{-15})\) and is a monotonically increasing function of time.

We conclude that current CMB data does not allow for a precise reconstruction of the inflationary power spectrum and the corresponding inflaton dynamics.  In particular, 
we find a statistical degeneracy amongst slow-roll and non-slow-roll models.  The span of the degeneracy envelope in Figure \ref{figure1} is
largely shaped by our inability to constrain \(r\), with models exhibiting the strongest
deviation from power-law also predicting values of \(r\) an order of magnitude larger than the fiducial
model.  Future CMB missions \cite{planck,inflationprobe} with the ability to detect the B-mode polarization signal
characteristic of gravitational radiation will put tighter constraints on \(r\)
\cite{Kinney:1998md,Verde:2005ff}, reducing or breaking the \(r\)-\(n_s\)
degeneracy and significantly improving primordial power spectrum resolution. 

\section{Conclusions}
We have investigated the ability of current data to constrain the form of the primordial power spectrum.
The ability of the data to resolve the power spectrum can be tested by how well it singles out a best-fit model.
Starting with the latest WMAP release, we generate a best-fit power law model consistent with the simplest models of
slow-roll inflation.  We test a wide variety of more complicated power spectra, and identify those that provide
equally good fits to the data relative to the best-fit model.  If the alternative spectra arise in an inflationary context, then we
can hope to identify both slow-roll and non-slow-roll models that are equally consistent with current data.

In order to test a wide array of inflationary power spectra, we turn to Monte Carlo reconstruction.  We combine the flow
formalism, which is a method of stochastic model generation, with a numerical integration of the mode equations of quantum
fluctuations.  This allows us to handle inflation models that yield spectra that are not well described by the
standard spectral parameterization.  For each spectrum thus generated, we fix the non-spectral parameters at their best-fit values
and calculate the likelihood of the model.  By fixing the non-spectral parameters, we are only sampling a subset of possible
spectra that might be degenerate with the best-fit power-law model, making this approach conservative.  We determine the
statistical significance of each model by obtaining the p-value calculated from the \(\chi^2_{eff}/{\rm d.o.f.}\), Eq.
(\ref{pval}).  

We generate an ensemble of power spectra to \(6^{th}\)-order in the flow space and select only those lying within \(\Delta p
\leq 0.01\) of the best-fit model.  A sampling of power spectra meeting this criterion are shown in Figure \ref{figure1}.  
The current CMB data provided by WMAP3 \cite{Spergel:2006hy} only reliably constrains the form of the power spectrum on
intermediate scales, \(0.01\ h{\rm Mpc}^{-1}\lesssim k \lesssim 0.1\ h{\rm Mpc}^{-1}\), with much variation on larger spatial
scales where cosmic-variance is the dominant source of error. 
By doing away with the spectral parameterization Eq (\ref{spec}), we also free ourselves from the constraints imposed on
the tensor-to-scalar ratio by SDSS data, the tightest bounds affecting models with a constant running of the
spectral index.  The flow method allows us to generate models with varied forms, including models with non-constant
running.  Such models can support a relatively large value of \(r_{0.002}\) by having a large running on these
scales that turns off on the intermediate scales probed by SDSS.  For example, we find spectra with \(r_{0.002} \sim
0.5\) that yield equally good fits to the WMAP+SDSS data as the best-fit power-law, a value which is well outside the \(r <
0.38\) \(95 \%\)-confidence limit for spectra with running.  By considering more general power spectra, one therefore
opens up regions of parameter space excluded in simpler models.   
The inflation models responsible for generating the strongly running spectra in this
study
are characterized by an initially fast-rolling inflaton.  Observable scales exit the horizon before the field slows to the
slow-roll attractor, yielding a large amplitude of gravitational waves and significant running.  We are able to conclude
that while slow-roll inflation models yield perhaps the  simplest explanation for the origin of large scale structure, {\it
fast-rolling} inflaton fields are equally suitable candidates from a strictly data-driven standpoint. Another interesting
approach is to expand the inflationary model space beyond the simplest class of canonical single field models. An example is the
DBI inflation scenario \cite{Silverstein:2003hf,Alishahiha:2004eh} analyzed using flow techniques by Peiris {\it et al.} \cite{Peiris:2007gz}, who find that such models are strongly constrained by the existing data.

Finally, we draw attention to the work of Lesgourgues and Valkenburg,
Ref. \cite{Lesgourgues:2007gp}.  In that analysis, the authors consider an enlarged inflationary parameter space consisting of the
coefficients of a Taylor expanded inflaton potential.  They perform a Bayesian parameter estimation analysis and obtain
confidence intervals on an ensemble of power spectra and their associated
inflaton potentials.  This choice of 
parameterization excludes models that deviate strongly from slow-roll, 
in particular, models of the type found in this analysis.  Aside from
this difference, we find good agreement
between our results, which are based on inferential statistics that are frequentist in nature, and the results of Ref.
\cite{Lesgourgues:2007gp}, derived using strictly Bayesian methods.  The bottom line is that current CMB data does not reliably constrain
the form of the power spectrum, and this conclusion can be reached from either a Bayesian or a frequentist approach.    

\section*{Acknowledgments}
This  research is
supported  in part  by  the National  Science  Foundation under  grant
NSF-PHY-0456777.

\end{document}